# Power Spectral Density-Based Resting-State EEG Classification of First-Episode Psychosis

Sadi Md. Redwan[1], Md Palash Uddin[2,3], Anwaar Ulhaq[4*] and Muhammad Imran Sharif[5]

[1]Department of Computer Science and Engineering, University of Rajshahi, Rajshahi 6205, Bangladesh (e-mail: sadi.redwan@ru.ac.bd)

[2]Department of Computer Science and Engineering, Hajee Mohammad Danesh Science and Technology University, Dinajpur 5200, Bangladesh (e-mail: palash_cse@hstu.ac.bd)

[3]School of Information Technology, Deakin University, Geelong, VIC 3220, Australia

[4]School of Computing. Mathematics and Engineering, Charles Sturt University, NSW, Australia (e-mail: aulhaq@csu.edu.au)

[5]COMSATS University Islamabad, Wah Campus, Punjab 47040, Pakistan (e-mail: mimraansharif@gmail.com)

[*]Corresponding author

**Abstract**

Historically, the analysis of stimulus-dependent time-frequency patterns has been the cornerstone of most electroencephalography (EEG) studies. The abnormal oscillations in high-frequency waves associated with psychotic disorders during sensory and cognitive tasks have been studied many times. However, any significant dissimilarity in the resting-state low-frequency bands is yet to be established. Spectral analysis of the alpha and delta band waves shows the effectiveness of stimulus-independent EEG in identifying the abnormal activity patterns of pathological brains. A generalized model incorporating multiple frequency bands should be more efficient in associating potential EEG biomarkers with First-Episode Psychosis (FEP), leading to an accurate diagnosis. We explore multiple machine-learning methods, including random-forest, support vector machine, and Gaussian Process Classifier (GPC), to demonstrate the practicality of resting-state Power Spectral Density (PSD) to distinguish patients of FEP from healthy controls. A comprehensive discussion of our preprocessing methods for PSD analysis and a detailed comparison of different models are included in this paper. The GPC model outperforms the other models with a specificity of 95.78% to show that PSD can be used as an effective feature extraction technique for analyzing and classifying resting-state EEG signals of psychiatric disorders.

**Keywords**: First-Episode Psychosis, EEG, PSD, GPC, Machine-Learning

## 1. Introduction

Psychosis is a symptom commonly associated with an extended array of neurological and psychiatric disorders, including schizophrenia spectrum (schizophreniform, schizoaffective, and paranoid schizophrenia). The first episode of psychosis in schizophrenia can be hard to distinguish from other forms of psychosis. An early diagnosis relies heavily on identifying trait markers of schizophrenia in First-Episode Psychosis (FEP/First-Episode Schizophrenia/FESz) patients. Electroencephalography (EEG) has been tremendously successful in the time-frequency analysis of neural activation patterns during different cognitive and behavioral assessments. Recent resting-state studies show that EEG can also be used to decode intrinsic brain activity in a task-negative state. Multiple studies involving spectral analysis support the alterations in resting-state delta/alpha activity in schizophrenia spectrum [1, 2, 36]. Several cortical alpha networks have been shown to be pathological in FEP patients in a recent magnetoencephalography (MEG) study [3]. Power Spectral Density (PSD) has been used in analyzing the alpha band Default Mode Network (DMN) in schizophrenia in another MEG analysis [4]. This raises the question of whether PSD can also be used for EEG analysis to identify FEP patients accurately.

In contemporary EEG and MEG studies, delta and alpha powers have been affiliated with attention and prolonged focus, signifying spontaneous resting-state brain activity. A more generalized model using multiple robust feature extraction techniques for highly accurate schizophrenia classification has also been proposed recently [5]. Several studies support the use of PSD as an effective EEG feature extraction method for machine-learning classification [37, 38]. In another study, researchers used PSD of multiple frequency bands along with fuzzy entropy and functional connectivity for Generalized Anxiety Disorder (GAD) classification with 97.83 (±0.4)% accuracy [6]. This signifies the potential utility of combining the spectral features of multiple bands for EEG classification of FEP. The core objective of this work is to combine the PSD of delta (0.5-4 Hz), theta (4-8 Hz), alpha (8-12 Hz), and sigma (12-16 Hz) bands of resting-state EEG for the machine learning approaches.

Machine learning models for EEG classification have been popularized with the success of Linear Discriminant Analysis (LDA), Support Vector Machine (SVM), and neural networks in multiple EEG paradigms. A random forest classifier has been proposed for the classification and analysis of mental states using single-channel EEG [7]. SVM has been successfully used in multiple sclerosis [8] and epilepsy detection [9]. Gaussian Process Classifier (GPC) has also been proposed for classifying mental states [10] and detecting neonatal seizures [11]. In this work, we analyze the effectiveness of multiple methods, namely random forest, SVM, and GPC, for classifying FEP patients and healthy controls based on the PSD of multiple EEG frequency bands. A medium-sized dataset of 28 controls and 44 patients has been balanced using borderline-SMOTE [12] for this work. With a very small number of parameters, the computationally efficient GPC has performed very well, with an accuracy of 95.51 (±1.74)% and a specificity of 95.78 (±3.3)%. The proposed framework sets a baseline for FEP and control classification using resting-state EEG, and we expect it to be improved upon in the future with more complex neural network models and multiple feature extraction techniques based on time-frequency analysis.

## 2. Materials and Methods

### 2.1 Electroencephalography (EEG)

EEG is a waveform representation of the (electrical) brain signals measured by the fluctuations of voltage induced by the neuronal ionic activity [13]. The effectiveness of EEG in decoding neurological and emotional states of the brain is attributed to the high temporal resolution of the signal [14] and our understanding of which frequency or pattern of the signal relates to a particular task, stimulus, or emotion. Several visual, auditory, and task-based stimuli have been developed over the years by researchers on account of EEG studies. These studies have eventually built the foundation of modern EEG-based emotion recognition, seizure detection, medical diagnosis, and Brain-Computer Interface (BCI) systems. In particular, EEG is currently established as the primary method for seizure detection [15]. Most publicly available EEG datasets are focused on diverse neural activation events of healthy and occasionally pathological brains. That being said, the publication of resting-state EEG studies and datasets has also increased in the past few years. Major depressive disorder [16], depression [17, 19], cognitive states [18], and multiple other psychiatric disorders [19] have been studied using resting-state EEG as of late, and some of them have been published as datasets. In addition to the MEG study of resting-state cortical alpha networks of FEP/FESz [3], Salisbury *et al.* also published the corresponding EEG datasets in 2022 [20, 21]. For our work, we use the *Resting Task 1* dataset, excluding the *Resting Task 2* samples of 10 subjects that are also present in the *Resting Task 1* dataset. The subject population consists of 72 subjects (28 controls and 44 patients). The demographic information of the subjects are presented in Table 1.

Table 1. Demographic information of the subject population.

| Group | N (male, female) | Average age (SD) | Ethnicity – White, Black, Asian, Mixed, Undisclosed |
|---|---|---|---|
| All subjects | 72 (46, 26) | 21.96 (4.66) | 46, 17, 5, 3, 1 |
| Control | 28 (16, 12) | 21.33 (3.88) | 21, 4, 3, 0, 0 |
| FEP | 44 (30, 14) | 22.36 (5.06) | 25, 13, 2, 3, 1 |

The dataset is obtained from OpenNeuro [22] (accession number: ds003944). It is available under the Creative Commons License (CC0). The phenotypic information is also included in the dataset. The cognitive and socio-economic assessments have been conducted using the MATRICS score and SES score respectively, and the negative effects of FEP are evident in the patient population.

**2.2 Preprocessing**

The initial step of every EEG study is preprocessing the data to reduce the effects of several unwanted artifacts. The EEG signals used in this work are obtained in a 5-minute period using a low-impedance 10-10 system 60-channel cap. Two additional electrooculogram (EOG) channels and an electrocardiogram (ECG) channel are also included in the data. EOG channels are particularly important as they capture the eye-blink artifacts that are also present in the EEG signals. Much work has been done to establish a correct method for EOG-related artifact removal based on Independent Component Analysis (ICA) and regression [23]. EEG signals also correlate with the ECG signal (heartbeat artifacts), which can be removed using ICA [24] and Signal-Space Projection (SSP).

ICA is a blind source separation (BSS) technique that has revolutionized signal separation from mixed signals and has been used in numerous EEG and fMRI studies over the years. With the success of a fast and efficient ICA implementation, fittingly named FastICA [25], it has become much easier to remove artifacts from EEG signals. In this work, FasstICA is used to remove both EOG and ECG artifacts separately. We apply temporal band-pass filtering of 0.5-35Hz before applying ICA to remove low-frequency drifts and high-frequency components that are not needed for this study. We extract 20 Independent Components (ICs) from all the channels to find out which components correspond to EOG and ECG artifacts and remove those components. The ICs for a sample subject are shown in Figure 1.

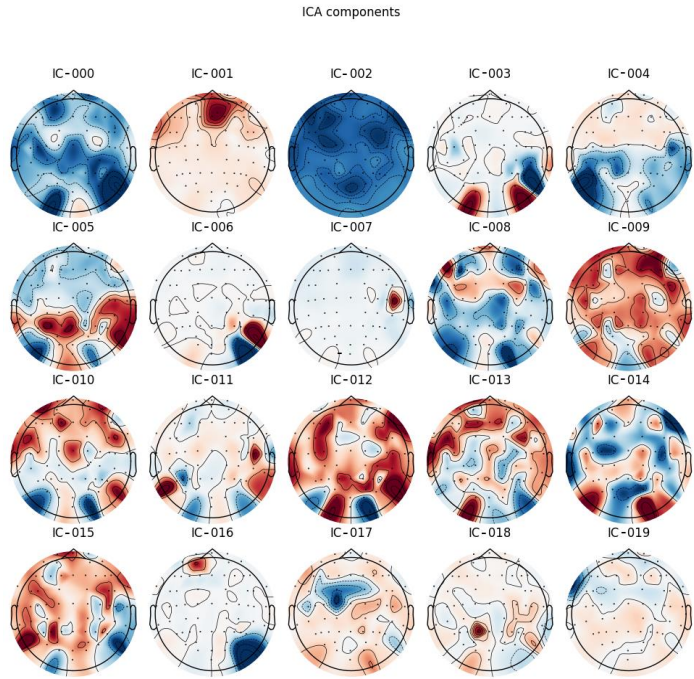

Figure 1. All 20 ICs for a subject. From a cursory glance, the IC-001 and IC-002 appear to be related to unwanted artifacts. IC-001 is close to the eyes, which indicates EOG-related potential, and IC-002 appears to be incoherent compared to the other ICs.

We identify ICs that are related to EOG artifacts by correcting the baseline (0.2 seconds interval) and averaging across the channels, as shown in Figure 2.

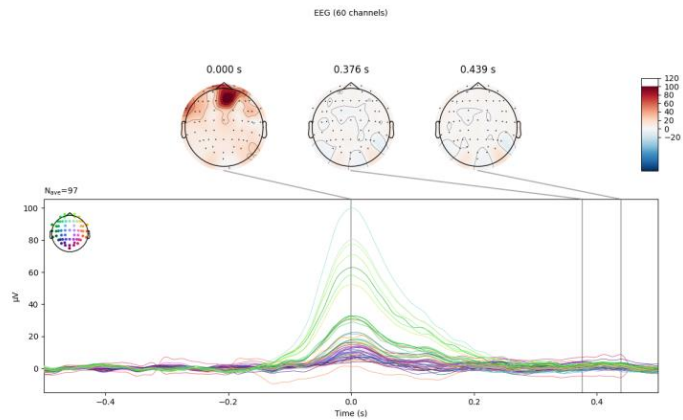

Figure 2. The ICs identified to be EOG-related IC (-0.5s–0.5s range, 1000 time points).

The ECG-related ICs are also identified using the same principle. Correlation is also applied to identify the heartbeat artifacts, since these artifacts do not affect each EEG electrode with the same potential due to the temporal properties of the ECG signal. Figure 3 shows the ICs that correlate to the ECG signal, and Figure 4 shows the effect of EOG and ECG-related artifact removal.

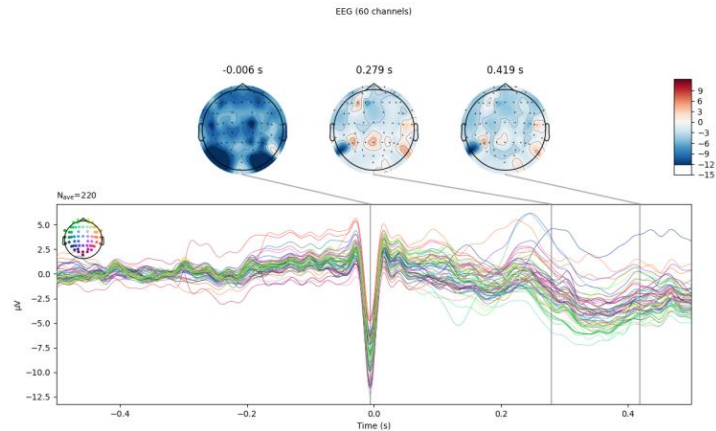

Figure 3. IC(s) identified to be ECG-related IC (-0.5s–0.5s range, 1000 time points).

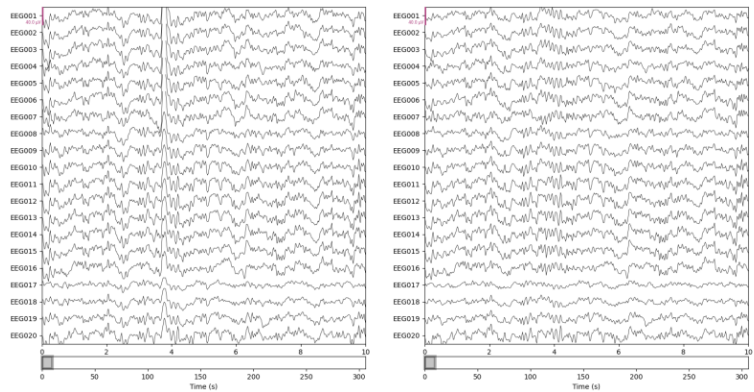

Figure 4. Effect of artifact removal. The original signals are shown in the left panel, and the processed signals are in the right panel. 20 out of 60 channels are shown with 0.5-16 Hz bandpass filtering in a 10s window; EOG artifacts are visible at ~4s timestamp in the left panel.

## 2.3 Cross-Spectral Density (CSD)

Before proceeding to the feature extraction step, we verify sensor-to-sensor coherence by calculating the CSD of the channels to justify using spectral features for further analysis. CSD compares two signals by measuring the spectral power distribution. There are different ways this can be achieved, for instance, using the Morlet wavelet (CWT/wavelet decomposition) and Short-Time Fourier Transform (STFT). We decompose every signal into time-frequency components using the Morlet wavelet to calculate the spectral correlation of the signals. For each frequency band, eight equidistant values (frequency scales) are specified from lower-bound to

upper-bound. The wavelet power spectrum can be defined as

$$(WPS)_x(\tau, s) = |W_x(\tau, s)|^2, \tag{1}$$

where $W_x$ is the wavelet transform and $\tau, s$ represent the position of the wavelet in the time and frequency domain, respectively [26]. The Morlet wavelet is given by

$$\psi(x) = exp\left(-\frac{x^2}{2}\right)cos(5x). \tag{2}$$

By combining the correlation between the power spectrums for each pair of signals, we eventually get a 60×60 matrix for all 60 channels. The average CSD matrices for a sample subject across different frequency bands are presented in Figure 5.

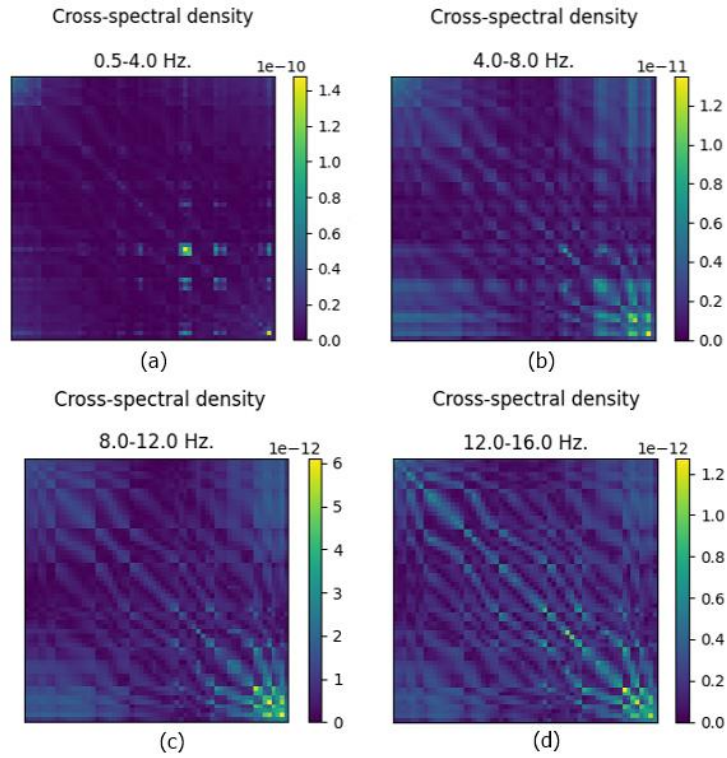

Figure 5. CSD analysis of a single subject. (a) delta, (b) theta, (c) alpha, and (d) sigma CSD matrices denote coherence across channel signals.

## 2.4 Power Spectral Density (PSD)

PSD is an effective method to differentiate between noise and features in a signal by making a spectral representation of the power distribution of its frequency components. We use Thomson's multitier spectral estimation [27] method to compute PSD. This method starts by calculating a periodogram for each of the first $K \approx 2NW$ Discrete Prolate Spheroidal Sequences (DPSS/Slepian tapers) [28] and then averaging these periodograms. Figure 6 shows the power spectra of a sample subject's preprocessed EEG data in $\mu V^2/Hz$ (decibels).

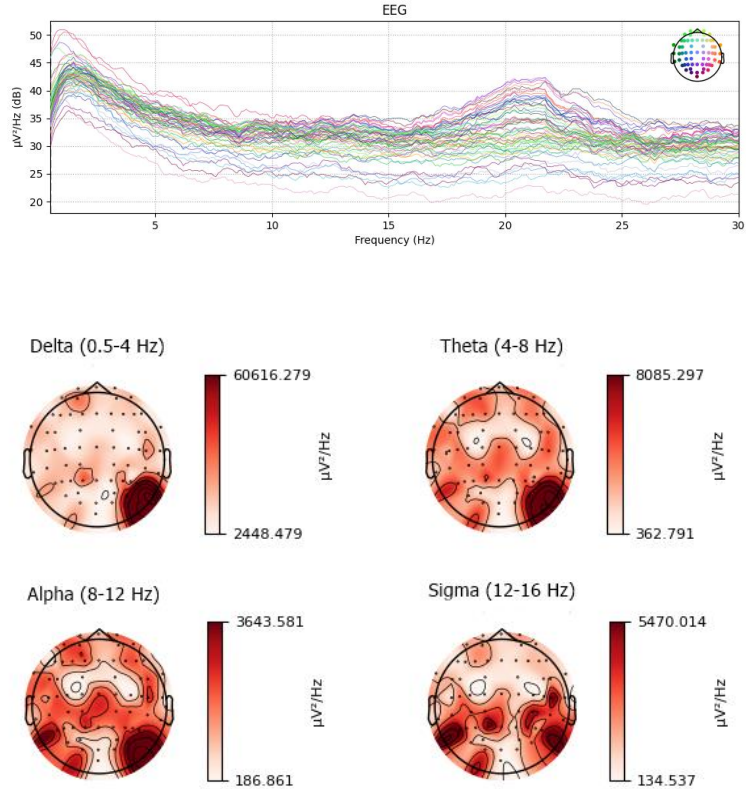

Figure 6. Power spectral representation of EEG data. Each frequency band shows the characteristic PSD of the signal.

We divide the data into 30s segments and compute four PSD bands for each subject. The four bands are then combined for the classification step.

## 2.5 Random Forest

Random forest is a tree-based ensemble learning technique [29] that has been used many times in different classification tasks. The core idea of a random forest classifier is to combine multiple decision trees using an ensemble (bagging) mechanism. The prediction of the random forest is given by the averaged prediction of the decision trees combined with the extremely-randomized method [30]. A random forest of 200 decision trees with a maximum depth of 30 per tree is used in this work to classify PSD feature vectors. Figure 7 presents a simple diagram of the random forest classifier.

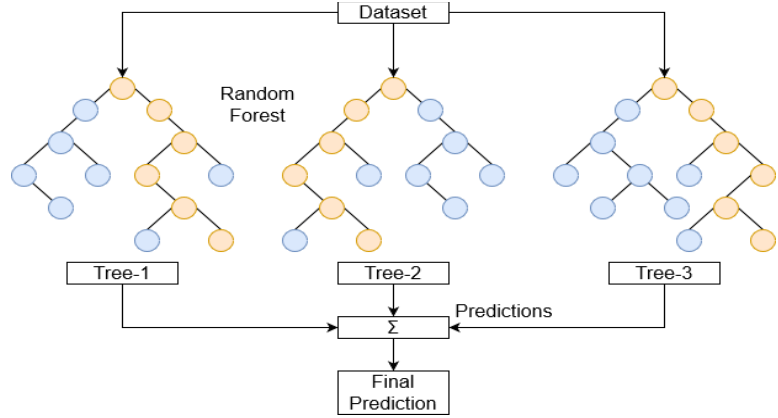

Figure 7. Random forest classifier architecture for binary classification.

## 2.6 Gaussian Process Classifier (GPC)

The GPC for binary classification is based on Laplace approximation [31]. With the joint probability $p(y)p(x|y)$ derived from Bayes' theorem, where $y$ denotes the class label, the marginal likelihood $p(y|X)$ is given by

$$p(y|X) = \int p(y|f)p(f|X)df = \int exp(\Psi(f))df. \tag{3}$$

Using a Taylor expansion of $\Psi(f)$ the approximation $q(y|X)$ to the marginal likelihood is derived as follows.

$$p(y|X) \simeq q(y|X) = exp\left(\Psi(\hat{f})\right) \int exp\left(-\frac{1}{2}(f-\hat{f})^T A(f-\hat{f})\right) df. \tag{4}$$

An approximation to the log marginal likelihood is derived by analyzing this Gaussian integral.

$$log\, q(y|X,\theta) = -\frac{1}{2}\hat{f}^T K^{-1}\hat{f} + log\, p(y|\hat{f}) - \frac{1}{2}log\,|B|, \tag{5}$$

where

$$|B| = |K|.|K^{-1} + W| = |I_n + W^{\frac{1}{2}}KW^{\frac{1}{2}}|, \tag{6}$$

and $\theta$ is a vector of hyperparameters of the covariance function.

We use a stationary covariance function, Radial Basis Function (RBF), as the Gaussian process kernel. With $r = \|x - x_i\|$ and a specified shape parameter $\varepsilon$, the Gaussian RBF is given as follows while the schematic working procedure of GPC is illustrated in Figure 8.

$$\varphi(r) = exp(-(\varepsilon r)^2) \tag{7}$$

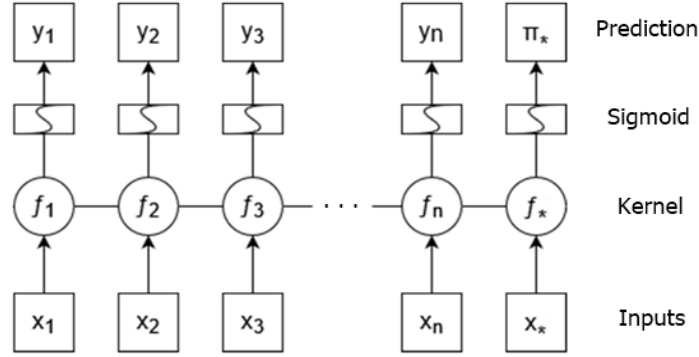

Figure 8. GPC architecture for binary classification.

## 2.7 Support Vector Machine (SVM)

Support vector machines (SVMs) are widely used for classification because they build a linear decision surface from a very large feature space to which input vectors are mapped non-linearly [32]. Based on the properties of the optimal hyperplane (feature map), the SVM algorithm can be classified into linearly separable, linearly inseparable, and non-linearly separable. For non-linear feature mapping, a kernel function is used to map the inputs implicitly. Similar to the GPC, we use the Gaussian RBF as the kernel function for our SVM model. For Gaussian RBF, $\varphi$ the kernel function can be written as

$$K(x_i, x_j) = \varphi(x_i).\varphi(x_j). \tag{8}$$

Then the vector to the hyperplane (weight) is given by

$$w = \sum_i \alpha_i y_i \varphi(x_i) \tag{9}$$

The SVM classifier minimizes the following expression to separate the input feature vectors with the parameter $\lambda > 0$, which denotes the tradeoff between the size and flexibility of the margin for classification while the basic architecture for non-linear SVM is shown in Figure 9.

$$\left[\frac{1}{n}\sum_{i=1}^{n} max(0, 1 - y_i(w^T x_i - b))\right] + \lambda |w|^2 \tag{10}$$

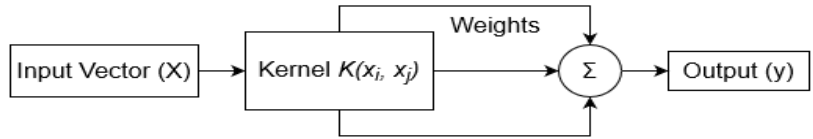

Figure 9. Non-linear SVM with Gaussian RBF kernel for binary classification.

## 3. Result and Discussion

The experiments were done using MATLAB R2022b and Python 3.10 in Microsoft Windows 11 (22H2) platform on an AMD Ryzen 7 3750H computer. The performance of each model is evaluated using 5-fold cross-validation. The final confusion matrix for each model is derived by taking the average of all confusion matrices, as shown in Figure 10.

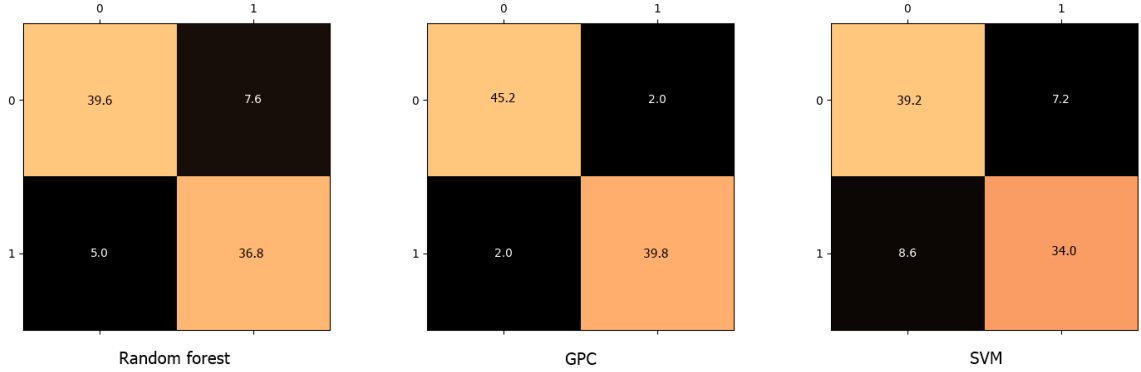

Figure 10. Average confusion matrices for test data.

We use precision, recall, and F1-score to evaluate the classification accuracy for each class. The mathematical expressions for precision, recall, and F1-score are as follows.

$$Precision = \frac{TP}{TP+FP}, \qquad (11)$$

$$Recall = \frac{TP}{TP+FN}, \qquad (12)$$

$$F1 = \frac{2 \times Precision \times Recall}{Precision+Recall}, \qquad (13)$$

where $TP$, $FP$, and $FN$ denote true-positive, false-positive, and false-negative predictions respectively. Specificity or true negative rate is defined as the recall of the negative class (control). The accuracy score, precision, recall, and F1 scores for the random forest, GPC, and SVM models are discussed in Table 2, Table 3, and Table 4, respectively.

Table 2. Classification report for the random forest model.

| Group | Precision (SD) | Recall (SD) | F1-score (SD) | Overall Accuracy (SD) |
|---|---|---|---|---|
| Control | 89.2 (±4.8) | 83.93 (±1.9) | 86.34 (±2.3) | 85.84 (±2.72) |
| FEP | 82.89 (±1.9) | 89.2 (±4.9) | 85.27 (±3.2) | |

Table 3. Classification report for the GPC model.

| Group | Precision (SD) | Recall (SD) | F1-score (SD) | Overall Accuracy (SD) |
|---|---|---|---|---|
| Control | **95.93** (±3.5) | **95.78** (±3.3) | **95.72** (±1.7) | **95.51 (±1.74)** |
| FEP | **95.56** (±3.5) | **95.3** (±3.1) | **95.26** (±1.8) | |

Table 4. Classification report for the SVM model.

| Group | Precision (SD) | Recall (SD) | F1-score (SD) | Overall Accuracy (SD) |
|---|---|---|---|---|
| Control | 82.49 (±3.6) | 84.69 (±4.2) | 83.45 (±2.2) | 82.25 (±2.18) |
| FEP | 82.47 (±3.4) | 79.45 (±5.3) | 80.75 (2.5) | |

With an accuracy of 95.51 (±1.74)% and specificity of 95.78 (±3.3)%, the GPC model has outperformed the other models (↑9.67% accuracy over random forest and ↑13.26% accuracy over SVM) and thus, decided as the best model for PSD-based classification of FEP vs. control. The proposed GPC model has a comparatively small number of parameters and can be considered a 'shallow' learning model. The high accuracy of GPC can be attributed to selecting a suitable covariance function for the input features. Other RBF kernels should also be considered for comparison. Deep recurrent neural network (RNN) models trained with time-frequency features, much like the recently proposed models for epilepsy classification, age prediction, and concussion classification [33, 34, 35], can hypothetically outperform this model. Another aspect that requires further analysis is the method for computing PSD. Future studies should also consider Welch's method for computing PSD to compare with the results of the DPSS method. Combining the CSD features with the PSD features can also provide insight into which electrode signals have the most significant impact on classification. This work can also be extended further for a spectrum-wide analysis of the schizophrenia spectrum.

## 4. Conclusion

In this study, we have evaluated the use of machine learning methods for the classification of patients with first-episode psychosis (FEP) and healthy controls based on the Power Spectral Density (PSD) of resting-state EEG. We have reviewed various feature engineering techniques and machine learning models to demonstrate that FEP patients can be accurately detected utilizing resting-state EEG. In addition, we have demonstrated that low-to-medium frequency (delta-to-sigma band) waves are pathological in FEP patients and can differentiate patients from healthy persons with the same degree of accuracy as task/event-related high-frequency waves. PSD is shown to be a reliable characteristic for the effective classification of FEP using machine learning. We conclude that resting-state EEG studies can lead to an accurate diagnosis of FEP/FESz and other psychiatric disorders and should be regarded as equally essential as stimulus-based EEG studies.

**Data Availability**

The denoised and preprocessed data used in this work is available at https://zenodo.org/record/7315010 while the original *EEG: First Episode Psychosis vs. Control Resting Task 1* dataset is available at doi:10.18112/openneuro.ds003944.v1.0.1.